\documentclass[manuscript]{aastex}

\usepackage{graphicx}
\usepackage{rotating}
\usepackage{dcolumn}
\usepackage{natbib}

\newcommand{\rsun}{R_{\Sun}}
\newcommand{\mydeg}{^{\circ}}

\newcommand{\rchi}{$\chi^2_{\nu}$ }
\newcommand{\rchiNS}{$\chi^2_{\nu}$}
\newcommand{\kms}{km s$^{-1}$ }
\newcommand{\kmsNS}{km s$^{-1}$}

\begin{document}

\title{Using ForeCAT Deflections and Rotations to Constrain the Early Evolution of CMEs}

\author{C. Kay}
\affil{Astronomy Department, Boston University, Boston, MA 02215}
\affil{now at Solar Physics Laboratory, NASA Goddard Space Flight Center, Greenbelt, MD, 20771}
\email{christina.d.kay@nasa.gov}

\author{M. Opher}
\affil{Astronomy Department, Boston University, Boston, MA 02215}

\author{R. C. Colaninno}
\affil{Space Science Division, Naval Research Laboratory, Washington DC 20375}

\and

\author{A. Vourlidas}
\affil{The Johns Hopkins University Applied Physics Laboratory, Laurel, MD 20723}

\begin{abstract}
To accurately predict the space weather effects of coronal mass ejection (CME) impacts at Earth one must know if and when a CME will impact Earth, and the CME parameters upon impact.  \citet{Kay15} presents Forecasting a CME's Altered Trajectory (ForeCAT), a model for CME deflections based on the magnetic forces from the background solar magnetic field.  Knowing the deflection and rotation of a CME enables prediction of Earth impacts, and the CME orientation upon impact.  We first reconstruct the positions of the 2008 April 10 and the 2012 July 12 CMEs from the observations.  The first of these CMEs exhibits significant deflection and rotation (34$\mydeg$ deflection and 58$\mydeg$ rotation), while the second shows almost no deflection or rotation ($<$3$\mydeg$ each).  Using ForeCAT, we explore a range of initial parameters, such as the CME location and size, and find parameters that can successfully reproduce the behavior for each CME.  Additionally, since the deflection depends strongly on the behavior of a CME in the low corona \citep{Kay15, Kay15AM}, we are able to constrain the expansion and propagation of these CMEs in the low corona.
\end{abstract}


\section{Introduction}
Coronal mass ejections (CMEs) are one of the key drivers of space weather at Earth.  To account for their effects, one must be able to forecast if a CME will hit Earth, as well as the timing and the CME parameters upon impact.  Recent efforts have focused on predicting the arrival time of CMEs.  Typically, a CME is observed in a coronagraph field-of-view, and the interplanetary motion is simulated based upon the interaction of the CME with the background solar wind \citep{Zha14, Hes15, May15AT, Shi15}.  These models tend to yield errors of 6-12 hours.

CMEs frequently show significant deflections from a purely radial trajectory (e.g. \citet{Hil77, Mac86, Byr10} and \citet{Isa14}).  Deflection can cause a high latitude CME to move toward the equator and impact Earth \citep{Byr10} or potentially cause a CME to miss the Earth when impact was expected \citep{May15, Mos15}.  CMEs tend to deflect toward the Heliospheric Current Sheet (HCS) and away from coronal holes (CHs, \citet{Cre04, Kil09, Gop09}).  Accordingly, magnetic forces have become a popular explanation for the cause of CME deflections \citep{Gop09, Gui11, She11, Kay15}.  In the low corona these forces will ``channel'' CMEs towards local null points \citep{Mos15, Wan15} and deflect CMEs toward the HCS on global scales.

CMEs have also been observed to rotate in both observations \citep{Gre07, Vou11, Nie12, Nie13} and simulations \citep{Tor03, Fan04, Lyn09}.  The rotation changes the orientations of the CME's magnetic field.  Knowing the direction of the magnetic field is crucial for space weather forecasting as the strength of geomagnetic storms, as measured by $Dst$, tends to increase with increasing southward magnetic field \citep{Yur05, Gop08}.

To begin addressing the questions about CMEs relevant to space weather predictions, \citet{Kay13,Kay15} developed Forecasting a CME's Altered Trajectory (ForeCAT).  ForeCAT initially addressed CME deflections due to magnetic forces.  With recent advancements ForeCAT is capable of simulating CME rotation as well \citep{Kay15}.  ForeCAT reproduces the general trends in CME deflections - CMEs tend to deflect toward the HCS on global scales, although the magnetic structure in the low corona can also affect the direction of deflection \citep{Kay15}.  The magnetic forces responsible for ForeCAT's deflection and rotation decay rapidly with distance, which causes the majority of the deflection and rotation to occur near the Sun \citep{Kay15,Kay15AM}.  Beyond 10 $\rsun$ a CME's deflection and rotation can be well described by assuming it propagates with constant angular momentum \citep{Kay15AM}.  \citet{Kay15L} and \citet{Pis15} show that ForeCAT can be used to reproduce observations of individual CMEs.

The angular momentum obtained in the low corona determines the behavior of CMEs at farther distances \citep{Kay15AM}.  Therefore it is essential to accurately describe a CME's behavior in the low corona, in particular the radial propagation and expansion of the CME.  However, it is often difficult to distinguish between deflection, rotation, and expansion in the low corona \citep{Nie12, Nie13}, many reconstruction techniques assume radial propagation and self-similar expansion (e.g. \citet{The06}). Since the deflection depends strongly on the CME's behavior in the low corona \citep{Kay15}, it is possible to use ForeCAT to probe the validity of different expansion models in the low corona.

In this paper we compare ForeCAT results with two observed CMEs - the 2008 April 10 and the 2012 July 12 CMEs.  Both CMEs were observed by the coronagraphs onboard the twin Solar Terrestrial Relations Observatory (STEREO) satellites.  These CMEs exhibit very different behavior with the 2008 April 10 CME showing a large deflection and rotation and the 2012 July 12 CME showing almost no non-radial behavior.  For both cases we fit the ForeCAT results to our CME positions, reconstructed from the STEREO observations, above 2 $\rsun$ and use this fit to place constraints on the CME expansion in the low corona.  In section 2 we briefly describe ForeCAT, and in sections 3 and 4 we present our results for the 2008 April 10 and the 2012 July 12 CMEs, respectively.

\section{ForeCAT}
ForeCAT determines the deflection and rotation of a CME based on the magnetic tension and magnetic pressure gradients determined from the background solar magnetic field.  Originally ForeCAT determined the motion of a CME cross section within a two-dimensional ``deflection plane'' \citep{Kay13}.  With the most recent version of ForeCAT \citep{Kay15} the full flux-rope is represented by a torus that is free to deflect in three dimensions.  The differential deflection forces along the torus cause a rotation about the axis connecting the CME nose and the center of the Sun, which changes the CME's tilt.  While observations show that other rotations can occur, they are not currently included in ForeCAT.  ForeCAT also includes the effects of drag in the nonradial direction, which opposes the deflection motion.

Currently ForeCAT describes the CME propagation and expansion using simple analytical or empirical models.  The radial speed follows a three-phase propagation model, similar to that presented in \citet{Zha06}.  The CME initially rises at a constant speed, $v_{\mathrm{min}}$.  At some distance $r_{\mathrm{ga}}$ the CME begins accelerating rapidly until it reaches a final speed, $v_{\mathrm{f}}$, at some distance $r_{\mathrm{ap}}$.  \citet{Kay15} show that the deflection can be sensitive to the chosen propagation model parameters.  In this work we use the observations to put initial constraints on these values and further constrain them through reduced chi-squared parameter space testing.  Typically we assume self-similar expansion for generic CME simulations, as this form of expansion tends to occur beyond 5$\rsun$ \citep{Che96, Che97, Woo09, Mie11}. However, CMEs frequently expand faster than self-similar in the low corona \citep{Che00, Cre04, Pat10a, Pat10b}.  When comparing with specific observed cases, such as in this work, we use an empirical description of the CMEs angular width versus distance.  Often the expansion is difficult to observe in the low corona but we can constrain this initial behavior through parameter space testing.  For a more thorough description of ForeCAT see \citet{Kay15}.

\section{2010 April 08 CME}
\subsection{Observations}
At 3:30 UT on 8 April 2010 a CME erupted from AR 11060, which was located at N25$\mydeg$E16$\mydeg$ as viewed from Earth (177$\mydeg$ Carrington longitude).  The polarity inversion line (PIL) of this active region (AR) was tilted ~40$\mydeg$ north of the solar equator.  \citet{Su11, Su13} and \citet{Kli13} determine the evolution of the magnetic field of the filament, which evolves into the CME, and the surrounding AR.  \citet{Su11} find that the filament becomes unstable as the axial flux increases as a result of flux cancellation near the PIL of the AR. As the filament erupts it quickly becomes inclined nearly 45$\mydeg$ with respect to the solar equator, an effect that can be reproduced with MHD simulations \citep{Kli13,Su13}.  The CME begins propagating radially by the time it reaches the STEREO/COR2-A field-of-view (2.5-15 $\rsun$).  This event was associated with an EIT wave and coronal dimmings \citep{Liu10EIT}.  As the CME was Earth-directed, \citet{Dav11} use it to test arrival-time prediction models.

\subsection{Reconstructed Position}
\begin{figure}[!hbtp]
\includegraphics[width=6.5in, angle=0]{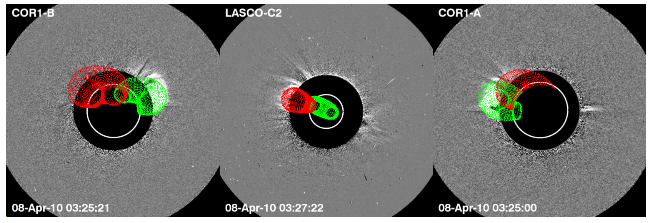}
\caption{GCS fits to the CME considered in this work (green) and a second CME (red) that occurred near the same time.  While the CMEs are spatially separated, they overlap in some coronagraph perspectives.}\label{fig:cgraphs}
\end{figure}

We determine the coronal trajectory of this CME using the Graduated Cylindrical Shell (GCS) model \citep{The06, The09}.  A separate CME without a strong EUV signature erupts more than 40$\mydeg$ westward of the CME considered in this work.   Their separation is sufficient that we can model the evolution of the 2010 April 8 CME without considering their interaction.  However, the two CMEs overlap in coronagraph images.  Figure \ref{fig:cgraphs} shows GCS fits to the CME of interest (green) and the other CME (red).  

From the GCS fits we determine the radial distance, latitude, longitude, tilt, and angular width of the CME versus distance. We reconstruct the position of the 2010 April 8 CME between 3:25 and 6:54 UT.  In this time the CME propagates radially from 1.8 $\rsun$ to 11.6 $\rsun$ while deflecting in latitude from -2$\mydeg$ to -7$\mydeg$, and remaining near 187$\mydeg$ longitude.  During this time the CME maintains a constant tilt of -23$\mydeg$ and a fixed angular width of 30$\mydeg$.  

\begin{figure}[!hbtp]
\includegraphics[width=6.5in, angle=0]{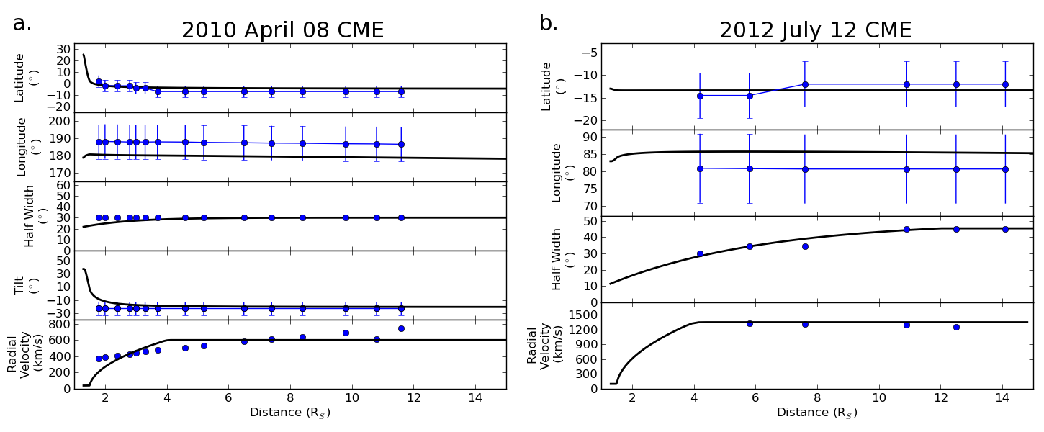}
\caption{Comparison of ForeCAT results (black line) with observations (blue circles) of the 2010 April 08 CME (Fig. \ref{fig:CPAs}(a)) and the 2012 July 12 CME (Fig. {fig:CPAs}(b)).}\label{fig:CPAs}
\end{figure}

Figure \ref{fig:CPAs} shows the latitude, longitude, half width, tilt, and radial speed versus radial distance.  To determine the radial speed we fit a quadratic polynomial to the radial distance as a function of time.  The radial speed is then determined as the derivative of this polynomial.  We assume the standard 5$\mydeg$ and 10$\mydeg$ uncertainties for the latitude and longitude from the GCS fit \citep{The09}.

Significant deflection must have occurred below 1.8 $\rsun$ as this CME originated at AR 11060, which is at 25$\mydeg$ latitude.  While only 5$\mydeg$ of deflection occurs between 1.8 and 11.6 $\rsun$, the total latitudinal deflection is closer to 30-35$\mydeg$.  This pattern of the largest deflection occurring close to the Sun matches the results of previous ForeCAT simulations \citep{Kay13, Kay15L, Kay15, Kay15AM, Pis15} as well as observed CME deflections \citep{Byr10, Gui11, Isa14}.  The AR longitude is within the 10$\mydeg$ error bars of the reconstructed CME position, so we cannot definitively confirm any westward deflection.  Additionally, we infer that this CME must have rotated as the reconstructed tilt differsmore than 60$\mydeg$ from the PIL of the AR.

\subsection{ForeCAT Results}
The left panel of Figure \ref{fig:CPAs} compares the best-fit ForeCAT results (black line, using the \citet{Kay15} version of ForeCAT) with the reconstructed CME latitude, longitude, angular width, tilt, and radial speed.  This best fit corresponds to the parameters listed in the first column of Table \ref{tab:bfparams}.  The free parameters we optimize are the initial latitude and longitude of the center of the CME, the tilt of the CME with respect to the solar equator, two shape parameters relating the height and cross-sectional width to the CME width ($A$ and $B$), as well as parameters related to models for the CME expansion ($m_{exp}$ and $b_{exp}$), propagation ($v_i$, $v_f$, $r_{ga}$, and $r_{ap}$), and mass evolution ($m_M$ and $b_M$.  We also consider the background drag coefficient $C_d$. 

We use an exponential increasing model for $\theta$, the angular width, 
\begin{equation}
\theta = b_{exp} + m_{exp}(1- \exp^{-R / R_{exp}})
\end{equation}

where $b_{exp}$ and $m_{exp}$ are free parameters related to the initial and final size of the CME, $R$ is the radial distance of the front of the CME, and $R_{exp}$ determines a scale height over which the expansion occurs. We use a $R_{exp}$ of 1.5 $\rsun$ for this case.  For the CME mass, $M_{CME}$, we use a linearly increasing model
\begin{equation}
M_{CME} = b_M + m_M R
\end{equation}
where $b_M$ and $m_M$ are coefficients given in Table \ref{tab:bfparams}.  

\begin{table}[!htp]
\begin{center}
\begin{tabular}{|l|c|c|}
\hline
Parameter & 2010 April 08 & 2012 July 12 \\
    \hline
Latitude ($\mydeg$)				& 24.9 	& -13.0\\
Longitude ($\mydeg$)				& 178.8 & 82.9\\
Tilt ($\mydeg$)					& 37. &  15\\
A = a/c						& 1		& 1 \\
B = b/c						& 0.1 	& 0.1\\
m$_{exp}$ ($\mydeg \rsun^{-1}$)			& 20 	& ---\\
b$_{exp}$ ($\mydeg$)				& 10 	& ---\\
v$_{i}$ (km s$^{-1}$)				&  40 	& 100\\
v$_{f}$ (km s$^{-1}$)				& 600 	& 1350\\
r$_{ga}$ ($\rsun$)				& 1.5 	& 1.5\\
r$_{ap}$ ($\rsun$)				& 4.0 	& 4.0\\
m$_{M}$ ($10^{14}$ g $\rsun^{-1}$)		& 0.2 	& ---\\
b$_{M}$ ($10^{14}$ g)				& 0.4 	& 10.\\
C$_d$						& 0.75 	& 1.0\\
\hline
\end{tabular}
\caption{Input parameters for the best fit cases.}
\label{tab:bfparams}
\end{center}
\end{table}

\begin{figure}[!hbtp]
\includegraphics[width=5.5in, angle=0]{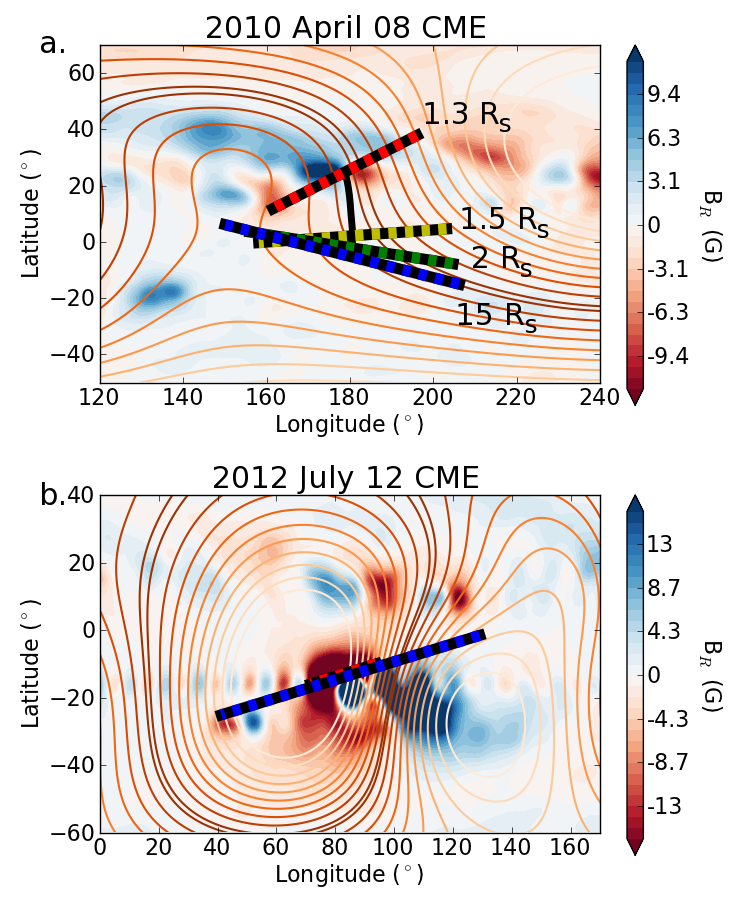}
\caption{Deflection and rotation of the 2010 April 08 CME (top) and the 2012 July 12 CME (bottom) and comparison with the magnetic background.  The color contours show the radial magentic field strength at the surface of the Sun, and the line contours indicate the total magnetic field strength farther out (dark lines indicating the weakest strength).  The dashed lines show the position and orientation of the CMEs at different distances.  For the 2012 July 12 CME we only show the initial and final position due to the minor deflection and rotation.}\label{fig:maps}
\end{figure}

The top panel of Figure \ref{fig:maps} compares the deflection and rotation of the 2010 April 08 CME with the solar magnetic background.  The color contours show the radial magnetic field at the surface of the Sun, revealing the location of the ARs.  The line contours indicated the total magnetic field strength farther out with the darkest lines corresponding to the weakest magnetic field, which shows the location of the Heliospheric Current Sheet.  The thick dashed lines indicate the orientation of the CME at several heights - 1.3 $\rsun$ (red, the beginning of the simulation), 1.5 $\rsun$ (yellow), 2 $\rsun$ (green), and 15 $\rsun$ (blue, the end of the simulation).  The solid black line shows the trajectory of the nose throughout the simulation.

This CME begins almost directly beneath the projection of the Heliospheric Current Sheet that forms at larger distances.  The CME exhibits a strong southward deflection (34$\mydeg$) and clockwise rotation (58$\mydeg$). The trajectory between 1.3 $\rsun$ and 1.5 $\rsun$ remains quasiparallel to the projection of the Heliospheric Current Sheet.  Initially the CME is oriented quasiperpendicular to the Heliospheric Current Sheet.  After the deflection and rotation, the edge of the CME closest to the Heliospheric Current Sheet lies nearly parallel to it.

\subsection{Variation with ForeCAT Parameters}
From the observations of the CME at coronagraph distances, we can loosely constrain the initial position and orientation of the CME and, to a lesser extent, the radial propagation and expansion models.  The mass and shape of the CME and background drag coefficient cannot be constrained from these observations alone.  For all free parameters, the behavior at distances below the coronagraph field-of-view is largely unconstrained.  To determine tighter constraint on these unknown paramters, as done in \citet{Kay15L} and \citet{Pis15}, we determine a best fit to the observations by sampling parameter space for the unknown ForeCAT input parameters. We determine the reduced chi-squared, \rchi,

\begin{equation}
\chi^2_{\nu} = \frac{1}{N-\nu-1} \Sigma \frac{(y_{obs} - y_{FC})^2}{\sigma^2_{obs}}
\end{equation}

where $N$ is the number of reconstructed positions, $\nu$ is the degrees of freedom, $y_{obs}$ are the observed positions, $y_{FC}$ are the ForeCAT positions, and $\sigma_{obs}$ is the uncertainty.  The $y$-values can correspond to either the latitude or longitude.  $y_{obs}$ and $y_{FC}$ must be compared at the same radial distance so we interpolate the ForeCAT results to the distance of the observations.  A \rchi near unity indicates a good fit, values significantly above or below correspond to a poor fit or overfitting the data.  In this work we assume an upper limit of 1.5 for a good fit according to the \rchi value.

\begin{figure}[!hbtp]
\includegraphics[width=6in]{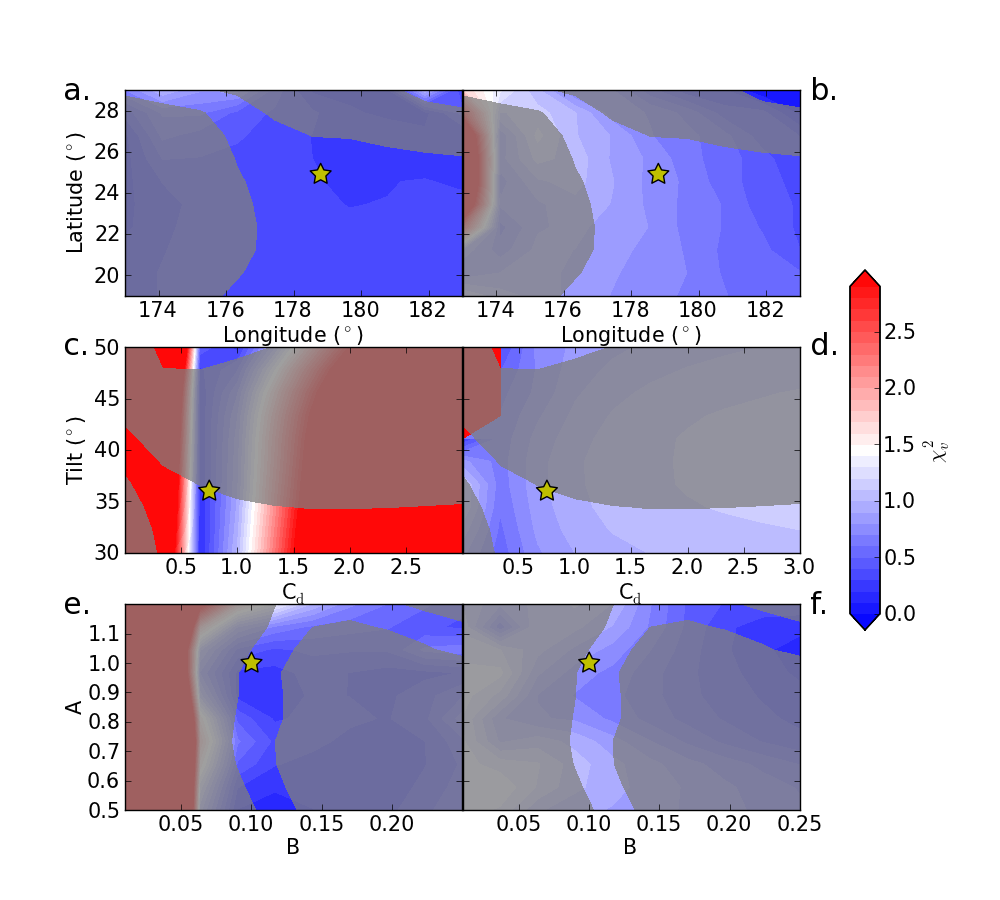}
\caption{Contours of \rchi for the initial CME latitude and longitude ((a) and (b)), the initial tilt and background drag coefficient, C$_{\mathrm{d}}$, ((c) and (d)), and the shape ratios $A$ and $B$ ((e) and (f)).  The left, and right columns correspond respectively to \rchi determined using the latitude or longitude.  The blue and red regions respectively correspond to good and poor fits to the observed deflection.  The shaded regions correspond to a final CME tilt that differs by more than 10$\mydeg$ from the reconstructed value, indicating a poor fit to the CME rotation. The yellow stars indicate the value of the best fit parameters.}\label{fig:1048c1}
\end{figure}

\begin{figure}[!hbtp]
\includegraphics[width=6.25in, angle=0]{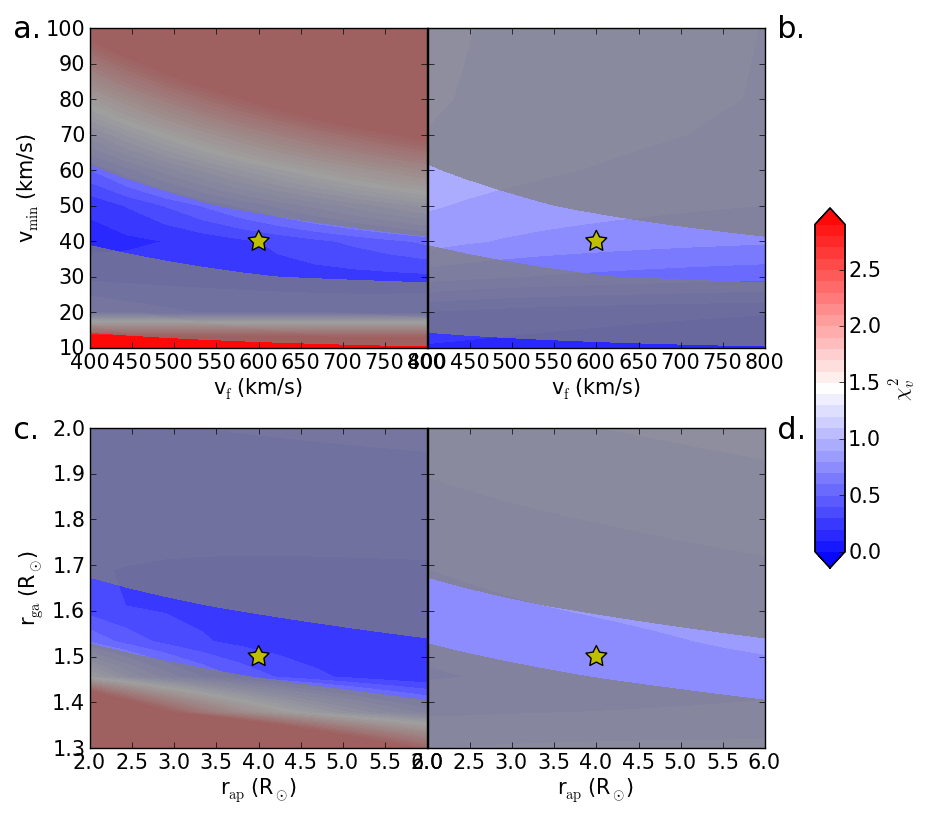}
\caption{Contours of \rchi for parameters related to the propagation model, analogous to Figure \ref{fig:1048c1}.  The top row shows the sensitivity to the initial and final CME speeds, $v_{\mathrm{min}}$ and $v_{\mathrm{f}}$, and the bottom row shows the transition distances from gradual rise to acceleration to constant propagation, $r_{\mathrm{ga}}$ and $r_{\mathrm{ap}}$.}\label{fig:1048c2}
\end{figure}

\begin{figure}[!hbtp]
\includegraphics[width=6.25in, angle=0]{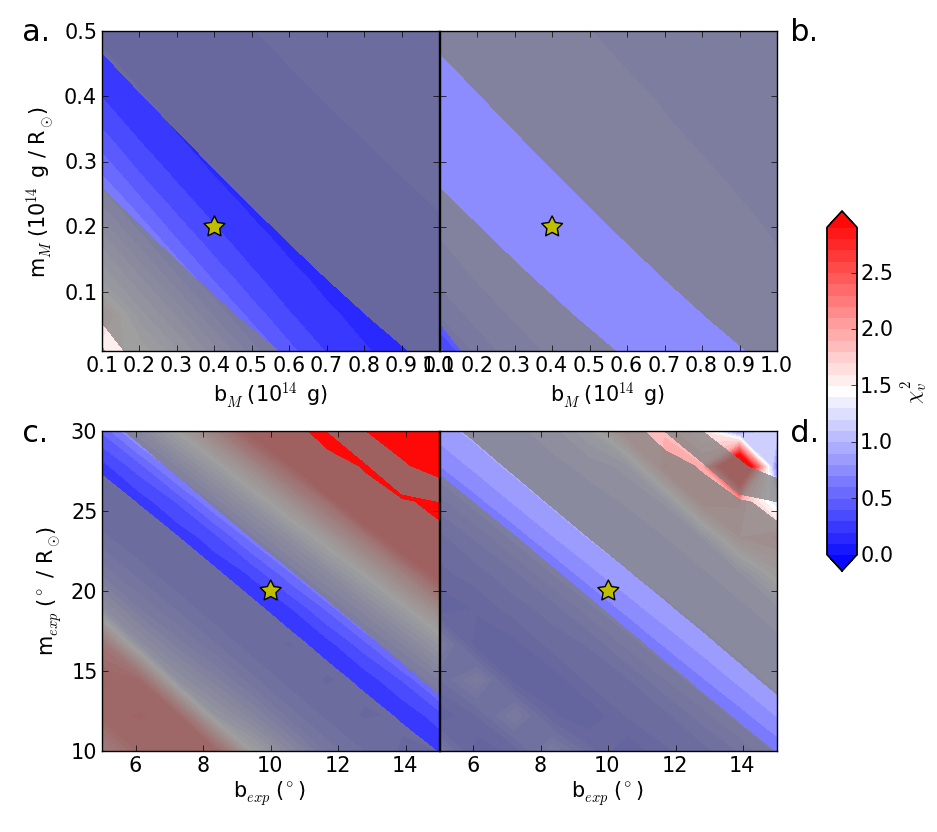}
\caption{Contours of \rchi for the linearly increasing CME mass (top, initial mass b$_{\mathrm{M}}$ and rate of increase m$_{\mathrm{m}}$) and exponential angular width models (bottom, initial width b$_{\mathrm{exp}}$ and increase in width m$_{\mathrm{exp}}$), analogous to Figure \ref{fig:1048c2} }\label{fig:1048c3}
\end{figure}

Figures \ref{fig:1048c1} through \ref{fig:1048c3} shows contours of \rchi for the initial parameters of ForeCAT.  The left columns show \rchi determined using only the reconstructed latitude, the right columns show \rchi using only the longitude.  The color contours are set so that white corresponds to a \rchi of 1.5, which is our cutoff for a good fit.  The blue regions corresponds to initial parameters that yield a good fit to the reconstructed trajectory. In each panel the values chosen for the best fit parameters are indicated with a yellow star.

We also consider the rotation of the CME when sampling parameter space.  For this CME, the majority of the rotation occurs below the coronagraph field-of-view.  As the reconstructed CME positions exhibit no rotation, we simply compare the final tilt of the modeled and observed CMEs rather than determining the \rchi using the tilt for each point in the CME trajectory.  The shaded region in Figure \ref{fig:1048c1} corresponds to the region of parameter space that yields a final CME tilt that differs by more than 10$\mydeg$ from the reconstructed value.

For many CMEs (excluding filament eruptions), the initial location can initially be loosley constrained to near a PIL of an AR.  We seek to further constrain the initial latitude and longitude using the deflection and rotation.  Figure \ref{fig:1048c1}(a) shows that the the latitudinal \rchi has little dependence on the initial CME position, but we see a small dependence on the longitudinal \rchi.  The CME rotation, however, is quite sensitive to the initial CME position, allowing us to restrict the initial positions to latitudes less than 26$\mydeg$ and longitudes greater than 177$\mydeg$.  For our best fit we use the initial position closest to the center of the PIL.

The middle row of Figure \ref{fig:1048c1} shows \rchi for different initial CME tilts and background solar wind drag coefficients.  We refer to the initial tilt as the tilt at the beginning of our simulation, which is analogous to the tilt at the beginning of a CME eruption.  Models show that kinked CMEs begin rotating early in an eruption \citep{Kli04, Tor03, Tor05}, so the initial tilt may differ from the first observed CME tilt.  Similar to the initial position, we begin loose constraints on the tilt from the observed PIL orientation, but the background drag coefficient is largely unknown, though assumed to be near unity.  Using the latitudinal \rchi between 0.5 and 1, but can place no constraints on the tilt.  The CME rotation is sensitive to the initial tilt, however, and we find two regions that reproduce the observed rotation.  We use an initial tilt of 37$\mydeg$, which is closest to the PIL orientation.

The bottom row of Figure \ref{fig:1048c1} shows \rchi for different values of the CME shape ratios $A$ and $B$, which are largely unknown from observations.  Using the latitudinal \rchi we can place a lower limit on $B$, but see little other sensitivity.  Once more we find that the rotation gives us tighter constraints on the free parameters.  The rotation is reproduced with CMEs with $B$ near 0.1 and any value of $A$.  Alternatively the observed rotation also occurs for CMEs with heights slighly larger than or comparable to their width (A $\geq$ ~1.1) and large cross-sectional widths (B $\geq$ ~0.1).

Figure \ref{fig:1048c2} shows contours of \rchi for parameters related to the radial propagation model (slow rise, rapid acceleration, constant propagation). Figure \ref{fig:1048c2}(a) and (b) show the sensitivity to the initial and final CME speed, $v_{\mathrm{min}}$ and $v_{\mathrm{f}}$.  The latitudinal \rchi allows us to put upper and lower bounds on the initial CME velocity.  The range of acceptable initial velocities decreases for faster final velocities.  The rotation  has similar behavior, but tighter constraints than the latitudinal \rchi.  If the initial speed is too low the CME spends too much time in the strong forces of the low corona, resulting in deflections and rotations larger than observations.  Conversley, a high initial speed yields insufficient deflection and rotation.  Since the majority of the deflection and rotation occurs before the CME reaches its final speed we are unable to constrain it.

Figure \ref{fig:1048c2}(c) and (d) show \rchi for variations in the radial distances at which the CME transitions from the gradual rise to acceleration phase and from the acceleration to the constant propagation phase, $r_{\mathrm{ga}}$ and $r_{\mathrm{ap}}$.  We find very similar behavior to the top panels of Figure \ref{fig:1048c2} - a strong sensitivity to the parameter related to the CME behaviour closest to the Sun, but little sensitivity to the parameter that determines the motion farther out. The longitudinal \rchi shows almost no sensitivity to these parameters.  The latitudinal \rchi and tilt depend strongly on the first transition distance, r$_{\mathrm{ga}}$, but not significantly on the second transition distance

Figure \ref{fig:1048c3} shows results for the parameters of the CME mass and angular width models.  The top row shows the dependence on the initial mass, $b_M$, and the rate at which the mass increases, $m_M$.  Figure \ref{fig:1048c3}(a) and (b) shows little variation with nearly the full range corresponding to \rchi less than 1.5.  By comparing the final tilt we can narrow the range of plausible parameters.  We find that higher values of $b_M$ require lower values of $m_M$.  We can eliminate sets of $b_M$ and $m_M$ where both parameters are near the high or low end of our considered region as these lead to either too little or too much rotation.

From the observations we know that the CME should have an angular width near 30$\mydeg$ by the time the nose reaches 2 $\rsun$.  However we have no measurements of the initial size of the CME, or the rate it increases, but we can constrain them from the \rchi in Figure \ref{fig:1048c3}.  The longitudinal \rchi shows little sensitivity to the expansion model parameters beyond a region of high \rchi for large values of $b_{exp}$ and $m_{exp}$.  We can, however, constrain the expansion using the latitudinal \rchi and eliminate the extreme sets where both parameters are either large or small.  As with the expansion we find that larger values of $b_{exp}$ require smaller values of $m_{exp}$.   The region of good tilts further constrains the good \rchi region, so that for any value of $b_{exp}$ we can constrain $m_{exp}$ within 5$\mydeg$.

\subsection{Discussion of Parameter Sensitivity}
For nearly all parameters we find that the longitude \rchi shows less variation than the latitudinal \rchi, which occurs for two reasons.  First, the longitude has twice the uncertainty of the latitude.  For the same difference between the simulated and reconstructed position the latitudinal \rchi will be four times the longitudinal \rchi.  Second, this CME shows significantly more latitudinal deflection than longitudinal deflection.  Parameters related to the CME's speed and mass tend to scale the total deflection, so the effects will be more noticeable in the direction where more deflection occurs. 

Every parameter, except for the drag coefficient, is more tightly constrained by the rotation than the deflection.  CME deflections and rotations result from both large-scale magnetic gradients, related to the orientation of coronal holes and the Heliospheric Current Sheet, and small-scale gradients related to the local structure of the AR \citep{Kay15, Kay15AM}.  For the 2010 April 8, CME we suggest that the deflection results predominantly from the large-scale gradients as we see little sensitivity to our free parameters.  These parameters change the position and size of the CME on relatively small scales, and the net effect is averaged out when integrated over the full CME.  The rotation, however, results from the distribution of these magnetic forces along the CME.  This causes the rotation to be much more sensitive to the initial parameters.  In other cases (see Section 4 or \citet{Kay15} and \citet{Kay15AM}) the local gradients can be important for deflection (or lack thereof), but we expect that even in such cases the rotation will still show greater sensitivity to our initial parameters.  Therefore, precisely measuring the rotation of CMEs will help us better constrain the early evolution of CMEs.

\section{2012 July 12 CME}
\subsection{Observations and Reconstruction}
On 12 July, 2012 a CME erupted from AR 11520 (S17$\mydeg$ W08$\mydeg$) accompanied by a X1.4 flare, which peaked at 16:45 UT.  The 2012 July 12 CME was much faster than the previously considered CME reaching a speed between 1,200 and 1,400 km s$^{-1}$ \citep{Hes14, Mos14, She14}.  \citet{Hes14} fit the CME flux rope and shock out to 80 $\rsun$ assuming a constant propagation direction of -8.9$\mydeg$ latitude and 0.3$\mydeg$ west of the Sun-Earth line (81.7$\mydeg$ Carrington longitude).  In situ observations show that this CME had a strong southward magnetic field \citep{Hes14, Mos14, She14}.  Previous studies of this event include a comparison with a MHD simulation \citep{She14}, and studies of the formation of the flux rope \citep{Che14} and reconnection during the eruption \citep{Dud14}.

As done for the 2010 April 08 CME, we reconstruct the CME's trajectory by fitting the GCS model to the coronagraph observations.  We determine the position of the 2012 July 12 between 16:54 and 18:24 UT, which corresponds to radial distances between 4.2 $\rsun$ and 14.1 $\rsun$. The latitude shows a small change from -14.5$\mydeg$ to -12.5$\mydeg$, and the longitude and tilt remain constant at 81$\mydeg$ and 28$\mydeg$, respectively.  The angular width increases from 30$\mydeg$ to 45$\mydeg$. 

The blue circles in Figure \ref{fig:CPAs} show the reconstructed position, longitude, width, and speed versus distance.  Again, we assume the standard uncertainties of 5$\mydeg$ and 10$\mydeg$ for latitude and longitude.  While the latitude and longitude do change slightly within our observed range, the values are consistent with no deflection from the original AR position due to our uncertainties.  Unlike the 2010 April 08 CME, we see that the angular width of the 2012 July 12 CME increases until about 10$\rsun$.  To mimic this slow, continued overexpansion we fit an exponential function of the form
\begin{equation}
\theta_{\mathrm{W}} (r) = \theta_{\mathrm{F}} (1 - \exp^{r/r_{\mathrm{W}}})
\end{equation}
where $\theta_{\mathrm{W}}$ is the angular half-width, $r$ is the radial distance, and $\theta_{\mathrm{F}}$ and $r_{\mathrm{W}}$ are free parameters representing the final CME width and the length scale over which the width varies.  We obtain a good fit to the observed width with $\theta_{\mathrm{F}}$ = 50$\mydeg$ and $r_{\mathrm{W}}$ = 5 $\rsun$.

\subsection{ForeCAT Results}
The right panel of Figure \ref{fig:CPAs} compares the best-fit ForeCAT results (black line) determined using the \rchi with the reconstructed position for the 2012 July 12 CME.  For this CME we do not include the effects of rotation.  We see no signature of rotation in the reconstructed position and the best-fit results change by less than 0.01$\mydeg$ in latitude and longitude and the CME rotates less than 1$\mydeg$.  The second column of Table \ref{tab:bfparams} contains the best fit parameters for this CME.  The ForeCAT results show that the CME does deflect from its original position, however it is a negligible amount - less than 0.5$\mydeg$ in latitude and 2.5$\mydeg$ in longitude.

The bottom panel of Figure \ref{fig:maps} compares the trajectory of this CME with the magnetic background.  The magnetic background is represented in the same format as the top panel.  For the 2012 July 12 CME we only show the initial and final position of the CME as there is little change in either the CME position or orientation.  Compared to the 2010 April 08 CME, this CME begins farther from the Heliospheric Current Sheet, and in an AR with stronger magnetic field, a position more favorable to large deflections and rotations.  The rapid propagation of this CME, however, nullifies these conditions so that the CME exhibits very little nonradial behavior.

\subsection{Variation with ForeCAT Parameters}
\begin{figure}[!hbtp]
\includegraphics[width=6in]{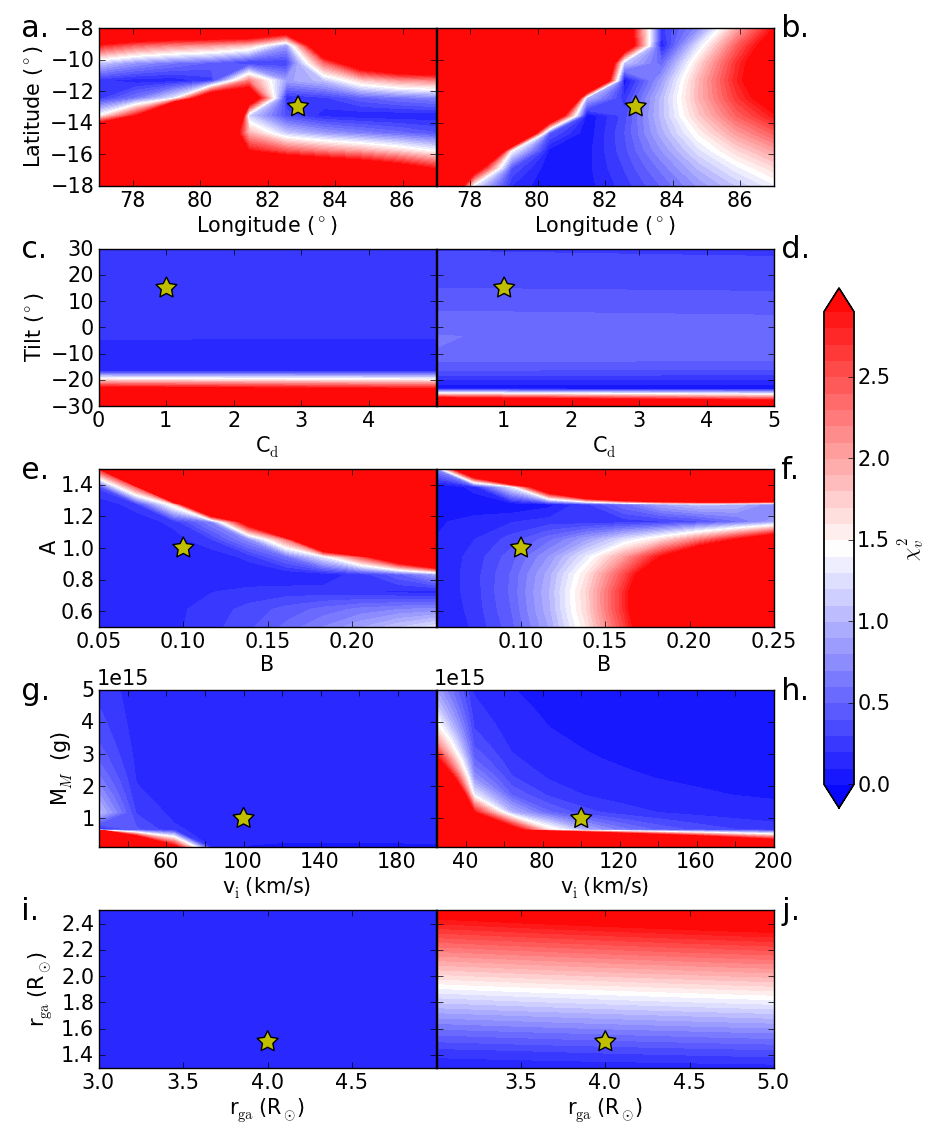}
\caption{Contours of \rchi for different ForeCAT input parameters for the 2012 July 12 CME.  The left column shows the latitudinal \rchi and the right column shows the longitudinal \rchiNS.  From top to bottom each row corresponds to the initial CME position (latitude and longitude), the initial tilt and solar drag coefficient, the CME shape parameters $A$ and $B$, the CME mass and initial CME velocity, and the distances of the transitions in the radial propagation model.}\label{fig:12712chi}
\end{figure}

Since we have fewer reconstructed points for this CME, and there is little difference, it is more difficult to constrain many of the CME parameters.  When a CME has significant deflection, the \rchi is very sensitive to parameters that affect the CME density, as determined from the CME mass and volume, as this causes the magnitude of the deflection to change.  Accordingly, since we see very little deflection, we do not consider a linearly increasing CME mass as it would be highly unconstrained.  Instead, we approximate the CME mass as constant.  Figure \ref{fig:12712chi} shows \rchi for different initial CME parameters, analogous to Figures \ref{fig:1048c1}-\ref{fig:1048c3}.  In Figure \ref{fig:12712chi} the top row shows \rchi determined using the reconstructed CME latitude, and the bottom row shows \rchi determined using the reconstructed CME longitude.

From the contours of \rchiNS, we find the tightest constraints on the initial CME latitude and longitude (Figure \ref{fig:12712chi}(a) and (b)).  Only a small range of initial latitudes and longitudes correspond to positions resulting in little to no deflection.  For all other positions near the AR PIL, the initial deflection forces are not balanced, leading to deflections greatly exceeding the observed values.  Combining the latitude and longitude \rchi gives an initial position between -14$\mydeg$ and -10$\mydeg$ latitude and longitude between 80$\mydeg$ and 84$\mydeg$.

The CME shape (Figure \ref{fig:12712chi}(e) and (f)) can also be constrained from the \rchi values.  The observed longitude cannot be reproduced with a cross-sectional width greater than 0.14 times the CME width ($B$ equals the ratio of these values), except for large values of $A$ that do not reproduce the observed latitude.  This limit on $B$ is comparable to the value found for the 2010 April 08 CME.  $A$ cannot exceed 1.4 for small values of $B$.  This upper limit is reduced to 0.9 for large values of $B$.

The rest of the initial parameters can, at best, be bounded on one side as \rchi is less than unity for much of parameter space.  From Figure \ref{fig:12712chi}(c) and (d) we determine the CME tilt less than 20$\mydeg$, but can put no constraints on the background drag coefficient.  The CME mass (Figure \ref{fig:12712chi}(g) and (h)) must be larger than 10$^{15}$ g, but any larger mass is acceptable as this serves to decrease the total deflection.  For more massive CMEs, any initial speed reproduces the results, but a mass as low as 10$^{15}$ g requires an initial speed above 80 km s$^{-1}$.  The latitudinal \rchi yields no constraints on the transition distance for the radial propagation model (Figure \ref{fig:12712chi}(i)), however, from the longitudinal \rchi we can constrain $r_{ga}$, the distance of the transition from the gradual rise phase to the acceleration phase to be less than 1.9 $\rsun$, which is slightly farther than the distance found for the 2010 April 08 CME.  Both the initial speed and the distance at which the CME begins accelerating determines how much time the CME spends in the low corona, which affects how long it is affected by the strong magnetic forces at these distances.

\section{Discussion and Conclusion}
In this work we consider two different CMEs - a very fast CME with little deflection, and a relatively slow CME that exhibits a significant deflection.  ForeCAT can reproduce the trajectory of both CMEs.  By determining the \rchi for hundreds of simulations sampling parameter space we can constrain some of the initial CME parameters that were previously unknown.  In addition to constraining parameters such as the initial CME position, orientation, and shape, we can also constrain the evolution of the CME width, speed, and mass in the low corona.  Both CMEs begin rapidly accelerating by 2 $\rsun$.  For the 2008 April 10 CME we can constrain the initial speed before this phase to between 30 \kms and 65\kmsNS.  Both CMEs overexpand in the low corona.  Both CMEs start with an initial angular width near 10$\mydeg$ before rapidly expanding.

In both this work and in \citet{Kay15L} and \citet{Pis15} we show that ForeCAT can reproduce the observed CME behavior when we provided with the observed propagation and expansion.  Predicting the occurrence of CME impacts at Earth, and the orientation of the CME's magnetic field is crucial for space weather forecasting.  Since it is computationally efficient ForeCAT could potentially be used to simulate a large range of CME parameters and determine the likelihood of CME impacts and the potential magnetic field orientations for any potential CME location, even before a CME occurs.  This, however, requires accurately simulating a CME's propagation and expansion from physical models rather than using the observed values after a CME has occurred.  Comparison with observed cases is already shedding light on the CME expansion and propagation in the low corona, and future work will focus on developing these aspects of ForeCAT.

\acknowledgements
C.K.'s research was supported by an appointment to the NASA Postdoctoral Program at NASA GSFC, administered by the Universities Space Research Association under contract with NASA.  A.V. acknowledges support from JHU/APL.  R.C.C. acknowledges the support of NASA contract S-136361-Y to NRL.  The SECCHI data are produced by an internation consortium of the NRL, LMSAL and NASA GSFC (USA), RAL and Univ. of Birmingham (UK), MPS (Germany), CSL (Belgium), IOTA and IAS (France).

\end{document}